\documentclass[manuscript]{aastex}

\shorttitle{Fine structure of flare ribbons}
\shortauthors{Sharykin and Kosovichev}

\begin{document}

\title{Fine structure of flare ribbons and evolution of electric currents}

\author{I.N. Sharykin\altaffilmark{1,2} and A.G. Kosovichev\altaffilmark{1,3,4,5}}
\affil{Big Bear Solar Observatory, New Jersey Institute of Technology,
    Big Bear City, CA 92314, U.S.A}

\altaffiltext{1}{Big Bear Solar Observatory}
\altaffiltext{2}{Space Research Institute of the Russian Academy of Science}
\altaffiltext{3}{Stanford University}
\altaffiltext{4}{NASA Ames Research Center}


\begin{abstract}
Emission of solar flares across the electromagnetic spectrum is often observed in the form of two expanding ribbons. The standard flare model explains the flare ribbons as footpoints of magnetic arcades, emitting due to interaction of energetic particles with the chromospheric plasma. However, the physics of this interaction and properties of the accelerated particles are still unknown. We present results of multiwavelength observations of C2.1 flare of August 15, 2013, observed with  New Solar Telescope (NST) of Big Bear Solar Observatory, Solar Dynamics Observatory (SDO), GOES and FERMI spacecraft. The observations reveal previously unresolved sub-arcsecond structure of the flare ribbons in regions of strong magnetic field consisting from numerous small-scale bright knots. We observe red-blue asymmetry of H$_{\alpha}$ flare ribbons with a width as small as $\sim$100 km. We discuss the relationship between the ribbons and vertical electric currents estimated from vector magnetograms, and show that Joule heating can be responsible for energization of H$_{\alpha}$ knots in the ribbons.

\end{abstract}


\section{Introduction}

Flare energy release in the lower solar atmosphere is usually observed in the form of two parallel ribbons radiating in a broad range of the electromagnetic spectrum. Emission from such ribbons can be stimulated by precipitating accelerated particles, Joule heating by electric currents, electron heat flux from a primary energy release site, or by a combination of these processes operating together. The standard (`thick-target') flare model \citep{Brown1971} assumes that the primary energy release occurs due to a magnetic reconnection process high in the corona, and that the low atmospheric phenomena represent a response to this energy release. However, there are evidences that the energy release may occur directly in lower regions of the solar atmosphere \citep{Fletcher2011}. In this case, the flare ribbons may also be generated by initial energy release in the lower solar atmosphere.

Knowledge of the fine structure of the flare ribbons and their links to the overlaying flare arcade is of great importance for the understanding of solar flares, because the dissipation rate of possible heating sources strongly depends on the value of the so-called `filling factor' of the emitting plasma.
The importance of the flare fine structure was realized long ago \citep[e.g.][]{Severnyi1957}. Recently,  \citet{Krucker2011} discussed implication of HINODE optical observations of flare ribbons for the `thick-target' model. Fluxes of nonthermal electrons per unit area, estimated in the framework of this model, can be explained only by assuming unexpectedly high density of accelerated electrons.
In such situation, smaller values of the filling factor can result in even more extreme flux densities of nonthermal particles.
In addition, loop-like structures (flare arcades) observed in the X-ray and microwave radio emissions clearly consist from multiple organized thin loops \citep[e.g.][]{Zimovets2013} which are also observed by TRACE in the UV range. \cite{Xu2012} discuss observations of flare kernels observed in the visible and near-infrared continua with characteristic sizes as small as $0.65^{\prime\prime}$. It appears that such fine structuring is an intrinsic property of the flaring plasma at all level of the solar atmosphere.

In this paper we present analysis of high-resolution observations of flare ribbons in the low atmosphere obtained with NST \citep{Goode2012} and simultaneous observations from three NASA spacecraft: SDO, GOES and FERMI. For the analysis we selected a flare of C2.1 GOES class, occurred on the August 15, 2013, approximately at 16:45:00 UT. The main criteria to select this event are its weakness (absence of extreme saturation of CCD) and the presence of large flare ribbons with relatively good seeing. The main task is to investigate fine structure of flare ribbons and their connection with properties of electric currents estimated from vector magnetic field measurements from SDO/HMI \citep{Centeno2014}.


\section{General description of the event}

The C2.1 flare occurred in active region NOAA 11818 near the disc center. The absence of coronal mass ejections (CME) and type II radio bursts indicates that the flare was non-eruptive. Also, there were no type III radio bursts, indicating on the absence of open magnetic field lines or large fluxes of accelerated electrons. Figure 1 presents a summary of the GOES and FERMI observations. The GOES data represent observations of soft X-ray emission integrated over the solar disc in two channels: 1-8 $\rm\AA$ and 0.5-4 $\rm\AA$ with time cadence of 2 seconds. These data (Figure 1-A) reveal a ``two-bump'' flare starting at $\sim$ 16:45 UT and ending at $\sim$ 18:17 UT (after this time the light curves are influenced by a flare from a different active region). The first GOES sub-flare in the long-wave channel peaked at 17:16 UT, and the second one had the maximum emission around 17:57 UT. In Figure 1-C we also show the flare emission measure and temperature, estimated following \cite{Thomas1985}. The temperature and emission measure profiles due to the background emission were removed.

The X-ray data from detector NaI-0 of the Gamma Burst Monitor (GBM) on the FERMI spacecraft are plotted in Figure 1-D with the time cadence of 0.25 seconds. We see that the shape of the GOES first peak is different from the FERMI count rate profile. The reason for this is that GBM is more sensitive to the hot plasma than the GOES detectors which mostly register a relatively low-temperature plasma. The hard X-ray (HXR) emission with photon energies greater than 27 keV (green curve in Fig.1-D) appears only for the second GOES peak at 17:51:56 UT; it correlates with the maxima of GOES X-ray flux derivatives for the both channels. The correlation between the flare HXR emission and the time derivative of soft X-ray (SXR) emission (Fig. 1-B) is known as the Neupert effect \citep{Neupert1969}. In the framework of the standard model, it corresponds to the initial plasma heating by accelerated electrons (HXR impulse) with subsequent chromospheric evaporation (SXR emission). In our case, the SXR time derivative starts increasing at $\sim$17:40 UT (second peak in Fig.1-B) at least 10 minutes earlier than the HXR peak (green curve in Fig.1-D), this may be due to additional energy release in the preheating phase. Moreover, during the first GOES sub-flare around 17:10:00 UT no HXR emission was observed. It appears that in this phase the plasma was heated to high temperature by some another mechanism different from the thick-target model. This event can be characterized as confined C2.1 flare with a minor signature of particle acceleration, but with separated extended ribbons.

\section{NST Data}

To investigate the spatial structure of the flare we use $H_{\alpha}$ data obtained with Visible Imaging Spectrometer (VIS) at the 1.6 m New Solar Telescope \citep{Goode2012}. The NST data catalog are available from http://www.bbso.njit.edu. The new system of adaptive optics AO-308 and large apperture make it possible to observe the Sun with high spatial resolution. The pixel size of the VIS images is about 0.029$^{\prime\prime}$, which is approximately 3 times smaller than the telescope diffraction limit $\lambda_{H\alpha}/D\approx 0.084^{\prime\prime}$. Final images are prepared by using a speckle reconstruction technique \citep{Woeger2008}. The time cadence between two subsequent line scans in 11 wavelength bands (6563 $\rm\AA$ $\pm$1.0, $\pm$0.8, $\pm$0.6, $\pm$0.4 and $\pm$0.2 $\rm\AA$) is 30 seconds. The NST observations started approximately at 17:10 UT; the data do not cover the begining of the first sub-flare, but include the whole second subflare.

In Figure 2 we present evolution of the flare H$_{\alpha}$ emission in comparison with the line-of-sight magnetic field observations from HMI. The time cadence of magnetograms is 45 seconds, and their spatial resolution is about 1$^{\prime\prime}$ (pixel size $\approx 0.5^{\prime\prime}$). To illustrate the structure of the flare emission in the lower solar atmosphere we selected three VIS channels: H$_{\alpha}$ red wing (+0.8 $\rm\AA$), H$_{\alpha}$ line center and H$_{\alpha}$ blue wing (-0.8$\rm\AA$). The contour lines display regions of strong magnetic field with magnitude $>1$ kG for comparison with bright features of the VIS data.

One of the interesting features observed by VIS is a dark filament (marked as F in Fig. 2) rooted near a small sunspot embedded into penumbra of the larger sunspot (marked as Y in the Fig. 2). Near the footpoint of the filament we see very bright sources in the H$_{\alpha}$ center, in addition to the two-ribbon flare structure. It is important to note that there is a strong incursion (shaped like a snake) of the opposite polarity magnetic field into the sunspot penumbra. This incursion ends near the rooted part of the filament F, and coincide with the accompanying bright H$\alpha$ sources.

There was no significant H$\alpha$ emission near the filament footpoint in both the blue and red line wings which reflect conditions in deeper layers of the solar atmosphere than the layer observed in the line core. Generally, images made in the H$_{\alpha}$ wings show weaker emission than in the line center. The strongest emission in the line wings is observed during the HXR pulse, in the form of two large-scale ribbons (marked in the Fig.2 as R1 and R2). Ribbon R1 has a diffuse structure and located in the region of relatively weak field ($<1$ kG), while the other ribbon R2 is very thin and located directly in the sunspot umbra, where the magnetic field reaches 2.3 kG. Since the very strong emission of the ribbons in the H$_{\alpha}$ center is observed long before the HXR impulse this means the initial plasma heating is likely not to be connected with the accelerated particles.

A remarkable property detected in the NST observations is the ultra-fine structure of the flare ribbons. To demonstrate this we present zoomed images of the flare ribbon R2 and Y-point near the root of the filament F in Figure 3 (bottom panels). The observed width of ribbon R2 in the red H$_{\alpha}$ wing sometimes is as small as 0.1$^{\prime\prime}$ ($\sim 70$ km). The ribbon has a clear fragmented structure in the form of tiny bright dots with the total area estimated as $\approx 3\times 10^{16}$ cm$^2$. The geometry of the flare region marked in Fig.2 as Y is very complex. It consists of many bright filaments and the overlaying darker filaments intersecting each other. Perhaps, this is a site of a very complex magnetic reconnection process involving penumbral flows and filament F.

In summary, the high-resolution NST data reveal superfine structuring of the chromospheric flare emission with a characteristic size of $\sim$100 km, organised in a very thin but long $\sim$10,000 km ribbon crossing the sunspot region. The other flare ribbon, located in a flare region, is spatially diffused but also displays small-scale structuring. In addition, the chromospheric emission of the flare ribbons starts long before the HXR impulse. The observed heating before the appearance of HXR emission cannot be explained by accelerated particles precipitating into the chromosphere, as assumed by the standard model. There must be additional heating mechanisms associated with the superfine organized structuring of the flare ribbons. Such fine structuring suggests that the heating mechanism may be related to Joule heating by chromospheric electric currents.

\section{SDO data, estimation of electric currents}

In this section we consider the evolution of vertical electric currents at the photosphere level. To calculate the vertical currents we use disambiguated HMI vector magnetic field data \citep{Centeno2014}. The time cadence is 720 seconds, and the spatial resolution is the same as for the line-of-site magnetograms (0.5$^{\prime\prime}$) per pixel. The vertical electric current density is calculated from the Ampere's law, $j_z=(\nabla\times\vec{B})_z/\mu_0 = (\partial B_x/\partial y - \partial B_x/\partial y)/\mu_0$.

Figure 4 displays a comparison of the electric current density with the flare images from the AIA \citep{Lemen2012} EUV data in two channels: He II 304 $\rm\AA$ and Fe XVIII 94 $\rm\AA$. The first channel has a peak in the temperature response function near $10^5$ K, while for another channel response goes up to 10 MK. The temporal and spatial AIA resolutions are correspondingly 12 seconds and $1.2^{\prime\prime}$ (with the CCD pixel size of $0.6^{\prime\prime}$).

The structure of electric currents in the flare region is very complicated. Numerous islands of electric current intensification are located near the polarity inversion line of the magnetic field and correspond to strong field regions. In the vicinity of the region marked as Y we observe the strongest electric currents during the whole flare. As we mentioned previously, this is a place of bright H$\alpha$ sources where filament F is rooted, and a place of intrusion of the opposite polarity magnetic field. This coincidence indicates that this region is probably heated by electric currents. The AIA 94 $\rm\AA$ images show the evolution of the hot plasma with temperatures comparable with estimations based on the GOES data (Fig.1-C). High coronal loops are rooted in the region of ribbons, and in the upper image we see a jet-like structure originating from near the Y-region.

Ribbon R1 observed in the 304 $\rm\AA$ channel does not coincide exactly with the electric currents observed near the line of magnetic polarity inversion, but the emission sources seen in the H$_{\alpha}$ line center partially correlate with these electric currents. In the region of ribbon R2 the enhanced electric currents correlate with a bright stripe seen  in both the AIA 304 $\rm\AA$ and VIS H$_{\alpha}$ images. In Figure 3 (right middle panel) we show the contour map of vertical electric currents ($j_z$), shows a spatial correlation of ribbon R2 with high values of $j_{z}$. Such correlation is more pronounced for the H $_{\alpha}$ emission sources near the Y-region.

To analyze the temporal dynamics of the electric currents in the flare region we select three boxes containing the Y-region, ribbons R1 and R2 (Fig. 5). We calculate mean value of $j_z$ in the boxes accounting only $j_{z}>max(j_z)/2$ to reduce contribution of the background. Errors are estimated as standard diviation of electric currents distribution in the quite Sun. In Figure 5 we compare the time profiles of the total vertical electric current in all three boxes with the evolution of the SXR emission according to the GOES observations. During the flares we see intensification of the vertical electric currents in the ribbons, but in the Y-region we observe local minimum. We also display the calculated time derivative of the total $B_z$ magnetic flux through all flare region, including boxes marked on the Fig. 5, as:

$$
\frac{dF}{dt} = d\left(\int B_{z}dS\right)/dt
$$

This value is proportional to the circuit electric current and can be considered as an indicator of the magnetic flux dynamics in terms of the electric field. It is one of the way to detect magnetic flux emergence or cancellation, events which could trigger a reconnection process. In Figure 5 we see that the main peak of $dF/dt$ corresponds to the HXR sub-flare. This probably means that the flare energy release is associated with the flux emergence.

In our study of the temporal dynamics of electric currents we use the HMI vector magnetic field data from 15:00 UT up to 21:00 UT. Figure 5 shows, that before the flares the electric currents are intensified and accompanied by a high rate of the magnetic flux change. But the HXR emission appears at $\sim$17:52 UT after the electric currents in ribbons R1 and R2 started increasing.

\section{Discussion}

The main result of the paper is the detection of extremely fine structuring of the H$_{\alpha}$ emission of flare ribbon located in the umbra of sunspot. The width of the ribbon emission is as small as $0.1^{\prime\prime}\approx 100$ km. In the case of the thin magnetic loops the density of electric current can reach high values as it depends on the value of filling factor. In the many works connected with flare studies filling factor is usually assumed to be a unity \citep[e.g.][]{SaintHilaire2005,Veronig2005}.

The electric current density estimated from the HMI data reaches $\sim 3\times 10^{10}$ À pixel$^{-1}$ or 0.4 A m$^{-2}$. Since the HMI pixel size is $0.5^{\prime\prime}\times 0.5^{\prime\prime}$, and the observed size of the ribbon knots is $0.1^{\prime\prime}\times 0.1^{\prime\prime}$. then the electric current density can be much greater. Assuming 5 knots in the HMI pixel (shaped like chain) the real density of the vertical electric current and field is 5 times larger.

In the regime of electric currents dissipation magnetic Reynolds number $Re_m=4\pi\sigma_{eff}L^2/(c^2\tau)\sim 1$, where $\sigma_{eff}$ is effective electric conductivity, $\tau$ is a characteristic time of electric current dissipation, and $L$ is a characteristic length scale which is taken as the size of observed H$_{\alpha}$ knots  $\approx 100$ km. From this formula assuming $\tau\sim 10^2-10^3$ seconds (duration of the HXR pulse and SXR flare) we obtain $\sigma_{eff}\sim 7\times 10^7-10^8$ s$^{-1}$. This value is somewhat lower than the theoretical estimates of electrical conductivity in the sunspot atmosphere, $\sim 10^8-10^9$ s$^{-1}$ \citep{Kopecky1966,Oster1968}. However, in the presence of magnetic field heating by electric currents in the partially ionized chromospheric plasma can be enhanced due to Pedersen conductivity. Also, the current dissipation may be enhanced due to turbulence.

The energy release rate of electric current can be estimated as $Q_{j}=j^2/\sigma_{eff}$. For the calculated values of $\sigma_{eff}$ and $j=0.4$ A m$^{-2}$ we have volumetric heating rate $Q_{j}\sim 20-200$ erg s$^{-1}$ cm$^{-3}$. To estimate radiation losses we use the assumption of optically thin plasma in the heated volume. As H$_{\alpha}$ emission corresponds to a typical temperature of $T=10^4$ K the value of radiation loss function is $f(T)\approx 10^{-23}$ erg s$^{-1}$ cm$^{3}$ \citep{Rosner1978}. Radiation energy losses are determined by formula $L_{rad}=n_en_Hf(T)$, where $n_e$ and $n_H$ are the electron and Hydrogen atom number densities. For $n_en_H\sim 10^{22}-10^{24}$ cm$^{-6}$ \citep{Avrett1981} $L_{rad}\sim 0.1-10$ erg s$^{-1}$ cm$^{-3}$ which is smaller than $Q_{j}$. Heat conduction losses estimated as $L_{cond}=4\times 10^{-6}T^{7/2}/L^{2}\approx 10^{-5}$ erg s$^{-1}$ cm$^{-3}$ are much less significant than $L_{rad}$. Thus, the observed emission in the H$_{\alpha}$ ribbon knots and plasma heating can be due to dissipation of electric currents. The largest part of $Q_j$ goes into the internal plasma energy $U_{th} = (n_H+n_e)k_bT$.

Together with the prominent fine structure in the red wing we observe similarly structured but weaker emission in the blue wing of H$_{\alpha}$ line. In a simplest way, such asymmetry can be explained by the Doppler shift due to plasma downflows.

\section{Results and conclusions}

The main observational results of the work are as the following:

\begin{enumerate}
\item The high-resolution NST observations reveal super-fine structuring of the flare ribbon located in the sunspot umbra. Numerous small bright knots as small as 0.1$^{\prime\prime}\approx 100$ km are organised in a $\sim 10^4$ km long regular thread-like structure. The second flare ribbon located in a plage region with weaker magnetic field is more diffuse, but also shows fine structuring. The ribbons are stronger in the red wing of H$_{\alpha}$ line than in the blue wing.
\item Vertical electric currents calculated from the HMI vector magnetograms spatially correlate with the observed H$_{\alpha}$ ribbons and emission sources. Temporal evolution of the vertical electric currents shows that a probable mechanism of the ribbon emission is heating by electric currents in the initial energy release phase.
\item The EUV AIA and H$_{\alpha}$ NST observations show enhanced emission long before the appearance of HXR emission, indicating active heating processes occured without accelerated high-energy particles.
\item The observed fine structuring of flare ribbons can be explained by the Joule dissipation if the electrical resistivity is enhanced in the partially ionized plasma of lower solar atmosphere, or by turbulent resistivity due to even smaller structuring.
\end{enumerate}

One of the most important open issues of this research is a very regular thread-like organisation of the superfine structuring of the observed ribbons. This large-scale organisation of the initial energy release needs further observational studies coupling with theoretical modelling.

Authors acknowledge the BBSO observing and technical team for their contribution and support. The work was partially supported by NASA grant NNX14AB70G and NJIT grant.


\begin{thebibliography}{18}
\expandafter\ifx\csname natexlab\endcsname\relax\def\natexlab#1{#1}\fi

\bibitem[{{Avrett}(1981)}]{Avrett1981}
{Avrett}, E.~H. 1981, in The Physics of Sunspots, ed. L.~E. {Cram} \& J.~H.
  {Thomas}, 235--255

\bibitem[{{Brown}(1971)}]{Brown1971}
{Brown}, J.~C. 1971, \solphys, 18, 489

\bibitem[{{Centeno} {et~al.}(2014){Centeno}, {Schou}, {Hayashi}, {Norton},
  {Hoeksema}, {Liu}, {Leka}, \& {Barnes}}]{Centeno2014}
{Centeno}, R., {Schou}, J., {Hayashi}, K., {Norton}, A., {Hoeksema}, J.~T.,
  {Liu}, Y., {Leka}, K.~D., \& {Barnes}, G. 2014, ArXiv e-prints

\bibitem[{{Fletcher} {et~al.}(2011){Fletcher}, {Dennis}, {Hudson}, {Krucker},
  {Phillips}, {Veronig}, {Battaglia}, {Bone}, {Caspi}, {Chen}, {Gallagher},
  {Grigis}, {Ji}, {Liu}, {Milligan}, \& {Temmer}}]{Fletcher2011}
{Fletcher}, L., {et~al.} 2011, \ssr, 159, 19

\bibitem[{{Goode} \& {Cao}(2012)}]{Goode2012}
{Goode}, P.~R., \& {Cao}, W. 2012, in Astronomical Society of the Pacific
  Conference Series, Vol. 463, Second ATST-EAST Meeting: Magnetic Fields from
  the Photosphere to the Corona., ed. T.~R. {Rimmele}, {et~al.}, 357

\bibitem[{{Kopeck{\'y}} \& {Kuklin}(1966)}]{Kopecky1966}
{Kopeck{\'y}}, M., \& {Kuklin}, G.~V.Retrieved 1~abstracts, s. w. n. . T. n. s.
  .~A. 1966, Bull. of the Astron. Inst. of Czechoslovakia, 17, 45

\bibitem[{{Krucker} {et~al.}(2011){Krucker}, {Hudson}, {Jeffrey}, {Battaglia},
  {Kontar}, {Benz}, {Csillaghy}, \& {Lin}}]{Krucker2011}
{Krucker}, S., {Hudson}, H.~S., {Jeffrey}, N.~L.~S., {Battaglia}, M., {Kontar},
  E.~P., {Benz}, A.~O., {Csillaghy}, A., \& {Lin}, R.~P. 2011, \apj, 739, 96

\bibitem[{{Lemen} {et~al.}(2012){Lemen}, {Title}, {Akin}, {Boerner}, {Chou},
  {Drake}, {Duncan}, {Edwards}, {Friedlaender}, {Heyman}, {Hurlburt}, {Katz},
  {Kushner}, {Levay}, {Lindgren}, {Mathur}, {McFeaters}, {Mitchell}, {Rehse},
  {Schrijver}, {Springer}, {Stern}, {Tarbell}, {Wuelser}, {Wolfson}, {Yanari},
  {Bookbinder}, {Cheimets}, {Caldwell}, {Deluca}, {Gates}, {Golub}, {Park},
  {Podgorski}, {Bush}, {Scherrer}, {Gummin}, {Smith}, {Auker}, {Jerram},
  {Pool}, {Soufli}, {Windt}, {Beardsley}, {Clapp}, {Lang}, \&
  {Waltham}}]{Lemen2012}
{Lemen}, J.~R., {et~al.} 2012, \solphys, 275, 17

\bibitem[{{Neupert} {et~al.}(1969){Neupert}, {White}, {Gates}, {Swartz}, \&
  {Young}}]{Neupert1969}
{Neupert}, W.~M., {White}, W.~A., {Gates}, W.~J., {Swartz}, M., \& {Young},
  R.~M. 1969, \solphys, 6, 183

\bibitem[{{Oster}(1968)}]{Oster1968}
{Oster}, L. 1968, \solphys, 3, 543

\bibitem[{{Rosner} {et~al.}(1978){Rosner}, {Tucker}, \& {Vaiana}}]{Rosner1978}
{Rosner}, R., {Tucker}, W.~H., \& {Vaiana}, G.~S. 1978, \apj, 220, 643

\bibitem[{{Saint-Hilaire} \& {Benz}(2005)}]{SaintHilaire2005}
{Saint-Hilaire}, P., \& {Benz}, A.~O. 2005, \aap, 435, 743

\bibitem[{{Severnyi}(1957)}]{Severnyi1957}
{Severnyi}, A.~B. 1957, \sovast, 1, 668

\bibitem[{{Thomas} {et~al.}(1985){Thomas}, {Crannell}, \& {Starr}}]{Thomas1985}
{Thomas}, R.~J., {Crannell}, C.~J., \& {Starr}, R. 1985, \solphys, 95, 323

\bibitem[{{Veronig} {et~al.}(2005){Veronig}, {Brown}, {Dennis}, {Schwartz},
  {Sui}, \& {Tolbert}}]{Veronig2005}
{Veronig}, A.~M., {Brown}, J.~C., {Dennis}, B.~R., {Schwartz}, R.~A., {Sui},
  L., \& {Tolbert}, A.~K. 2005, \apj, 621, 482

\bibitem[{{W{\"o}ger} {et~al.}(2008){W{\"o}ger}, {von der L{\"u}he}, \&
  {Reardon}}]{Woeger2008}
{W{\"o}ger}, F., {von der L{\"u}he}, O., \& {Reardon}, K. 2008, \aap, 488, 375

\bibitem[{{Xu} {et~al.}(2012){Xu}, {Cao}, {Jing}, \& {Wang}}]{Xu2012}
{Xu}, Y., {Cao}, W., {Jing}, J., \& {Wang}, H. 2012, \apjl, 750, L7

\bibitem[{{Zimovets} {et~al.}(2013){Zimovets}, {Kuznetsov}, \&
  {Struminsky}}]{Zimovets2013}
{Zimovets}, I.~V., {Kuznetsov}, S.~A., \& {Struminsky}, A.~B. 2013, Astron.
  Letters, 39, 267

\end{thebibliography}

\clearpage

\begin{figure}
\epsscale{.80}
\plotone{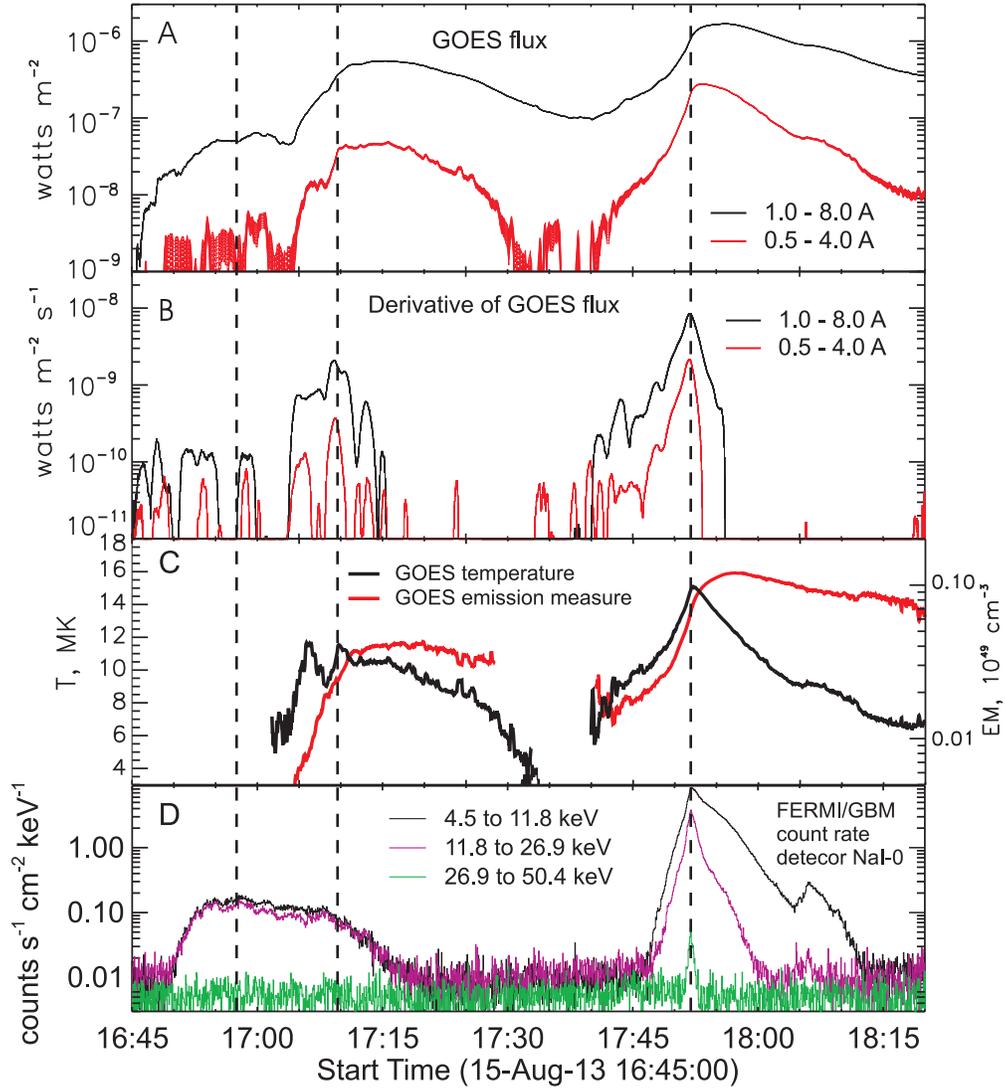}
\caption{Panel A: GOES X-ray fluxes in two channels: 1.0-8.0 $\rm\AA$ and 0.5-4.0 $\rm\AA$. Panel B: time derivatives of the GOES X-ray fluxes. Panel C: Temperature and emission measure derived from the GOES X-ray data. Panel D: FERMI X-ray count rates in three energy bands indicated in the plot.\label{fig1}}
\end{figure}
\clearpage
\begin{figure}
\epsscale{0.9}
\plotone{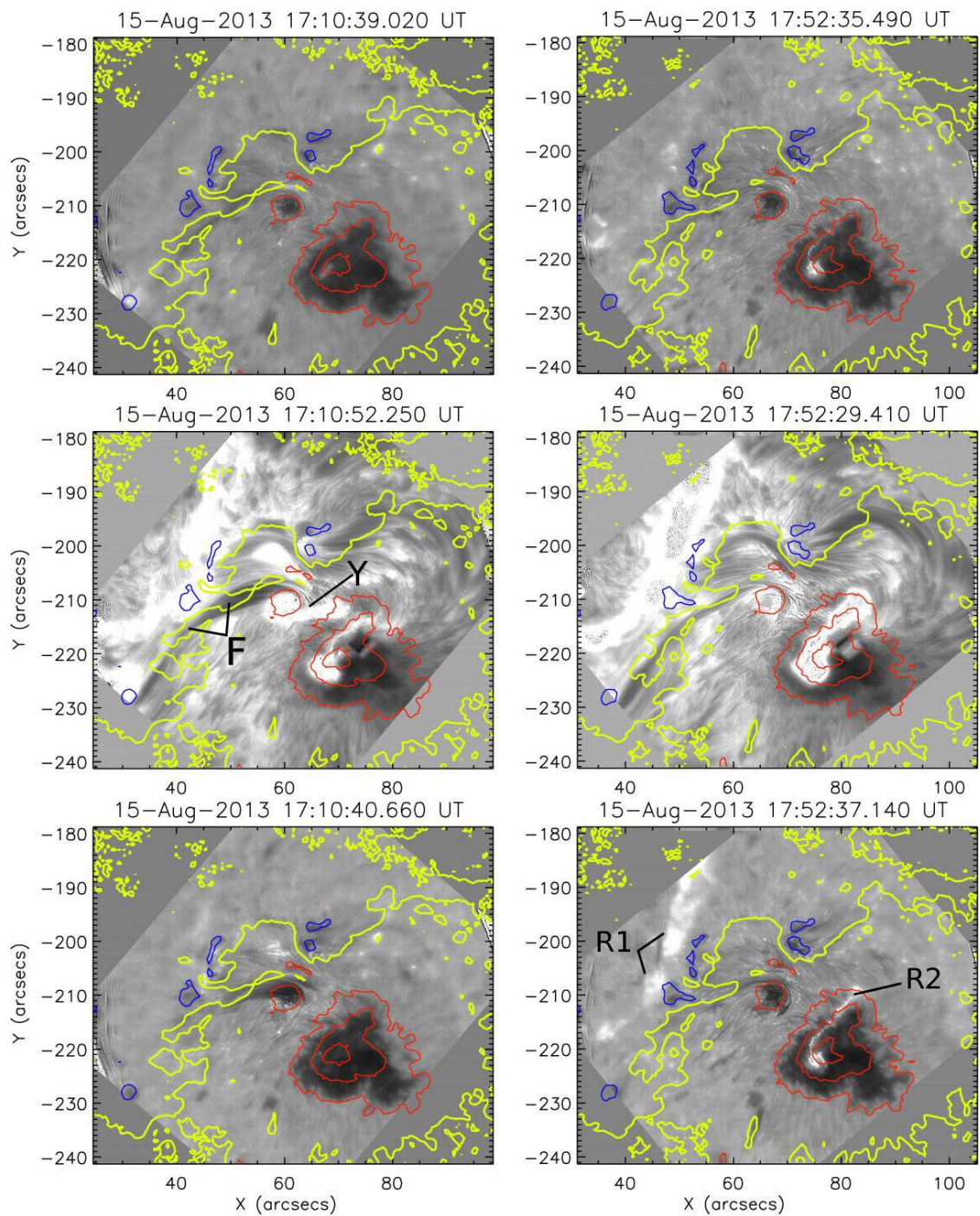}
\caption{NST observations made in: the blue wing of H$_{\alpha}$ line -0.8 $\rm\AA$ (top panels), the H$_{\alpha}$ center (middle panels) and the red wing of H$_{\alpha}$ line +0.8 $\rm\AA$ (bottom panels). Red and blue contours correspond to the positive and negative line-of-sight magnetic field with strength of 1, 1.5 and 2 kG. Green lines mark the polarity inversion line of the magnetic field. \label{fig2}}
\end{figure}

\clearpage
\clearpage
\begin{figure}
\epsscale{0.75}
\plotone{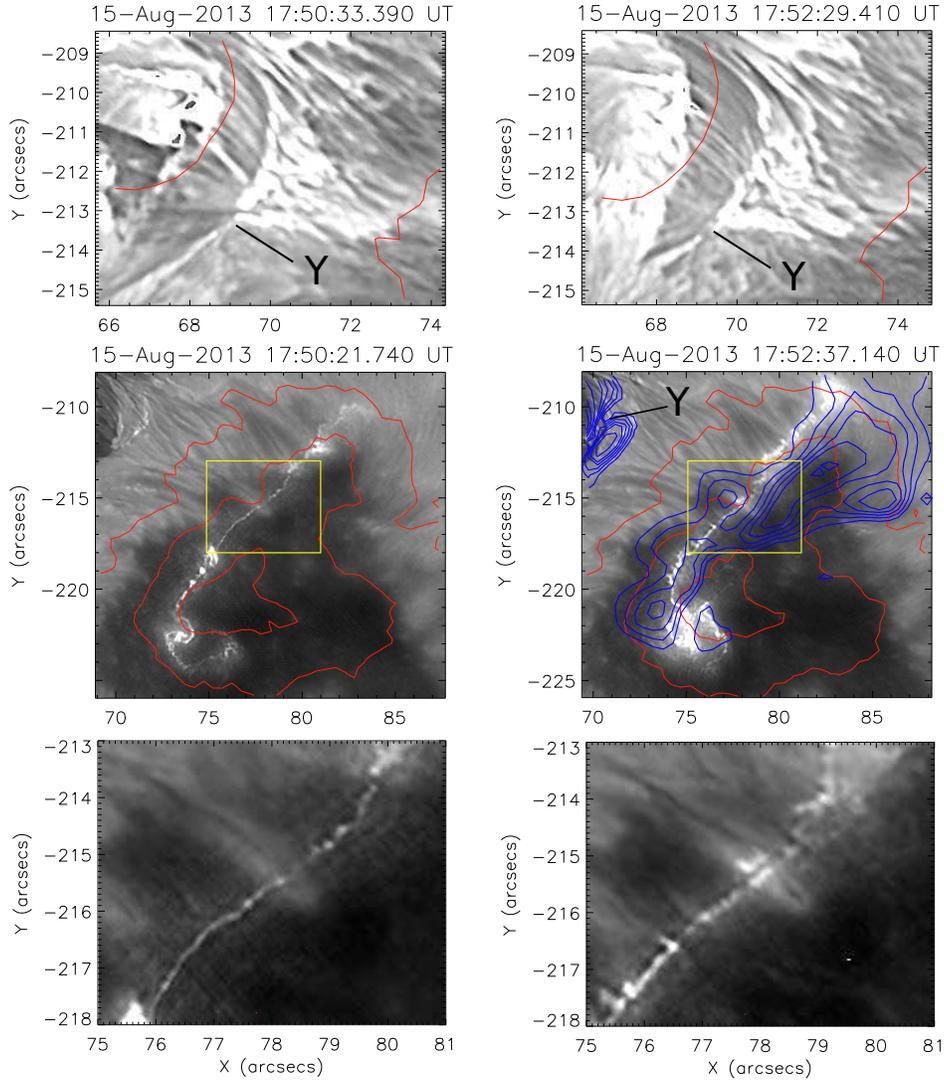}
\caption{Top panels: The structure of the Y region observed in the center of H$\alpha$ line. Middle panels: ribbon R2 in the red wing of  H$\alpha$ line +0.8 $\rm\AA$. Bottom panels: zoomed ribbon R2 in the box marked by yellow rectangles in the middle panels. Contours mark magnetic field with strength of 1, 1.5 and 2 kG.\textbf{ Blue contours on the right middle panel display vertical electric current with levels 10, 15, 20, 25, 30, 40, 50, 60, 70 and 90 \% of maximum.} \label{fig3}}
\end{figure}

\clearpage
\clearpage
\begin{figure}
\epsscale{0.9}
\plotone{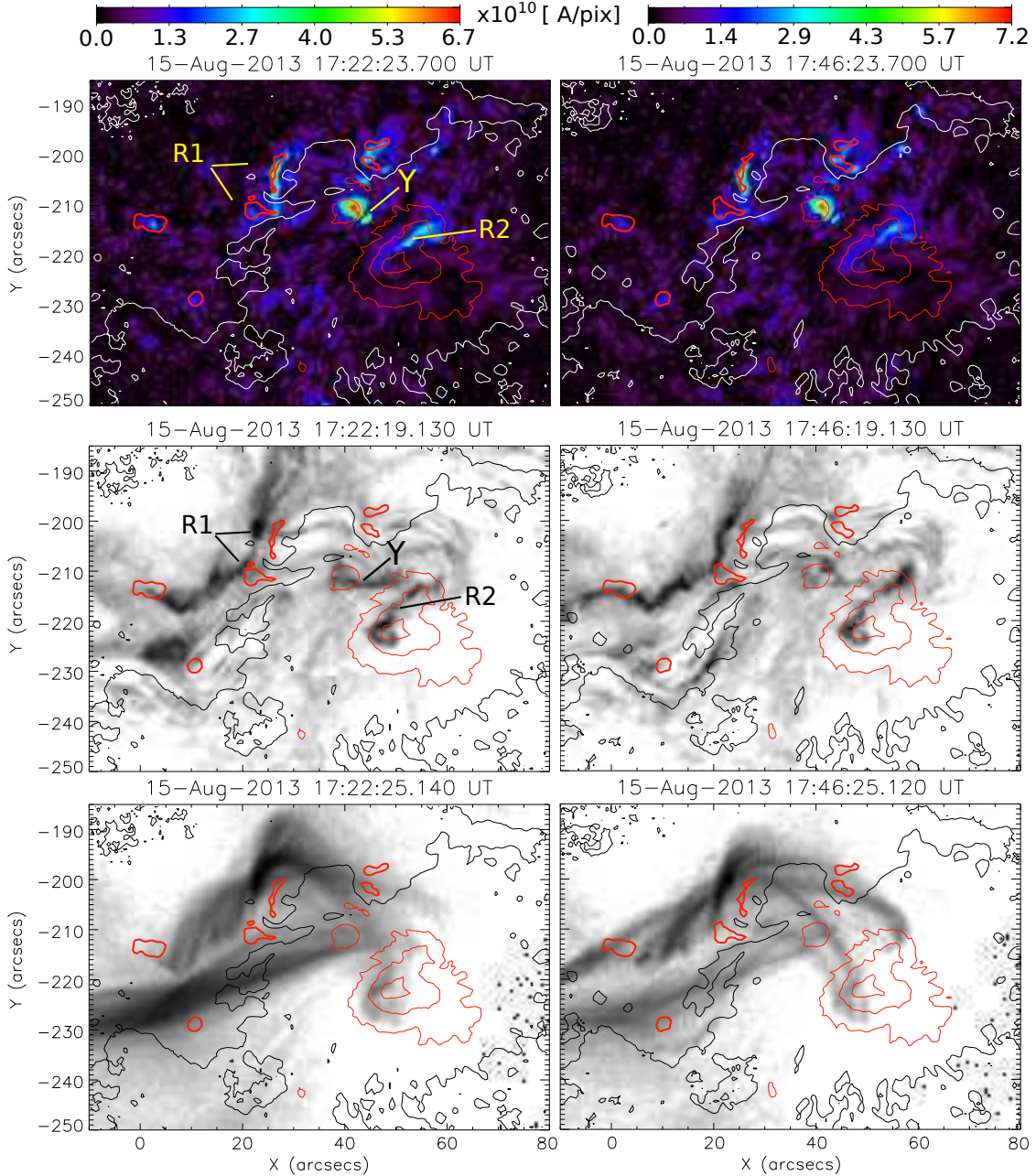}
\caption{Distributions of the vertical electric current density (top panels), and coronal emission in the AIA 304 $\rm\AA$ channel (middle panels), and in the 94 $\rm\AA$ channel (bottom panels). Overlayed contours on the images correspond to the absolute values of the line-of-sight magnetic field of 1, 1.5 and 2 kG. White and black contours mark the magnetic field polarity inversion line.\label{fig4}}
\end{figure}

\clearpage
\clearpage
\begin{figure}
\epsscale{1.0}
\plotone{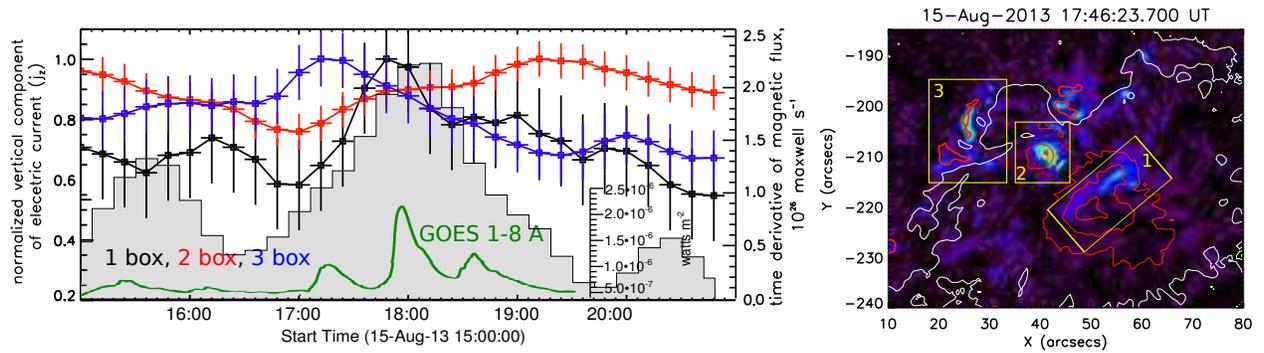}
\caption{Black, red and blue lines in left panels correspond to the mean electric current density from the boxes 1,2 and 3 marked by yellow rectangles on the right image. Grey histogram is the time derivative of the total magnetic flux in the flare region. Green curve shows the X-ray flux from the GOES long-wave channel.\label{fig5}}
\end{figure}

\clearpage
\end{document}